\documentclass[twocolumn,pre,reprint,showpacs,preprintnumbers,amsmath,amssymb,superscriptaddress]{revtex4-2}
\usepackage{graphicx}
\usepackage{dcolumn}
\usepackage{bm}
\usepackage{amsmath}
\usepackage{physics}
\usepackage{multirow}

\let\oldAA\AA
\renewcommand{\AA}{\text{\normalfont\oldAA}}

\usepackage{upgreek}
\usepackage{amsmath}
\usepackage{subfigure}
\usepackage{hyperref}
\usepackage[normalem]{ulem}
\hypersetup{
    colorlinks,
    citecolor=blue,
    filecolor=black,
    linkcolor=blue,
    urlcolor=blue
}
\usepackage{epstopdf}
 \usepackage{relsize}

\newcommand*{\rom}[1]{\expandafter\@slowromancap\romannumeral #1@}
\title{Report}

\setcitestyle{square}

\begin{document}


\title{Thermometry in dual quantum dot set-up with staircase ground state configuration}
\author{Sagnik Banerjee}

\affiliation{%
	Department of Electronics and Telecommunication Engineering, \\Jadavpur University,  Jadavpur-700032, India\\
}%
\author{Aniket Singha}


\affiliation{%
Department of Electronics and Electrical Communication Engineering, \\Indian Institute of Technology Kharagpur, Kharagpur-721302, India\\
}%






\begin{abstract}
We propose and investigate thermometry of a setup employing dual quantum dots with staircase ground state configuration. The stair-case  ground state configuration  actuates thermally controlled inelastic tunnelling, which translates into a temperature sensitive conductance, thereby inducing thermometry. The performance of the set-up is then analyzed employing  quantum master equation (QME) for such systems in the sequential tunnelling regime. In particular, it is demonstrated that the system performance, in terms of temperature sensitivity and efficiency, is maximum in the regime of low temperature, making such system suitable for cryogenic thermometry. The proposed set-up can pave the path towards realization of high performance cryogenic nano temperature sensors. \\
\end{abstract}
\maketitle
\section{Introduction}
Cryogenic thermometry has recently attracted significant attention with particular focus towards nano-scale systems. Electronic thermometry is actuated via temperature-induced control of electronic transport, which has manifested in the form of nano-scale thermoelectric engines \cite{thermoelectric_transport_at_mesoscopic_level_1,aniket_nonlocalheat,thermoelectric_transport_at_mesoscopic_level_2,heatengine1,heatengine2,heatengine3,heatengine4,heatengine5,aniket,heatengine6,bd1,bd2,bd3,aniket_heat1,aniket_heat2,heatengine7,heatengine8,heatengine9,heatengine10,heatengine11,heatengine12,heatengine13,heatengine14,heatengine15,heatengine16,heatengine17}, refrigerators \cite{thermalrefrigerator1,thermalrefrigerator2,aniket_nonlocalref,aniket_cool1,aniket_cool2,thermalrefrigerator3,thermalrefrigerator4,thermalrefrigerator5,thermalrefrigerator6,bd4,thermalrefrigerator7,thermalrefrigerator8}, transistors \cite{thermaltransistors1,thermaltransistors2,thermaltransistors3,thermaltransistors4,thermaltransistors5,thermaltransistors6,thermaltransistors7,thermaltransistors8} and rectifiers \cite{thermalrectifier1,thermalrectifier2,thermalrectifier3,thermalrectifier4,thermalrectifier5,thermalrectifier6}. Theoretical investigation of cryogenic nano-thermometers employing quantum dots and quantum point contacts has been of immense interest recently due to their applications in the temperature regime of a few Kelvin. Despite such technology being in their infant stages, tremendous effort is being geared to enhance the performance of such set-ups in chip level systems.\\
\indent In this paper, we propose and investigate cryogenic thermometry in a set-up employing dual quantum dots with staircase ground state configuration. The quantum dots are embedded in a nano-wire like structure. An equivalent set-up with multiple quantum dots was conceived earlier in literature as a means to optimize heat harvesting \cite{LijieLi}.  The energy difference between adjacent dot ground-states actuates inelastic process assisted tunneling, accompanied by energy absorption or emission. Thus, any change in the system temperature impacts the intensity of inelastic phenomena and thus induces a change in conductance. It is demonstrated that the thermometry performance in terms of temperature sensitivity and efficiency is optimal in the low-temperature regime, making such a set-up extremely suitable for cryogenic applications. \\
\indent This paper is organized as follows. In Sec \ref{design} we elaborate on the proposed system configuration. Next, in Sec \ref{analysis}, we discuss the electronic transport formulation employed to analyze the thermometry performance of the proposed set-up. Sec \ref{results} elaborates on the performance and operating regime of the proposed set-up. Finally, we conclude the paper briefly in Sec \ref{conclusion}
  \section{Proposed Device Structure}\label{design}
  \begin{figure*}
  	\centering
  	\includegraphics[width=1.05\textwidth]{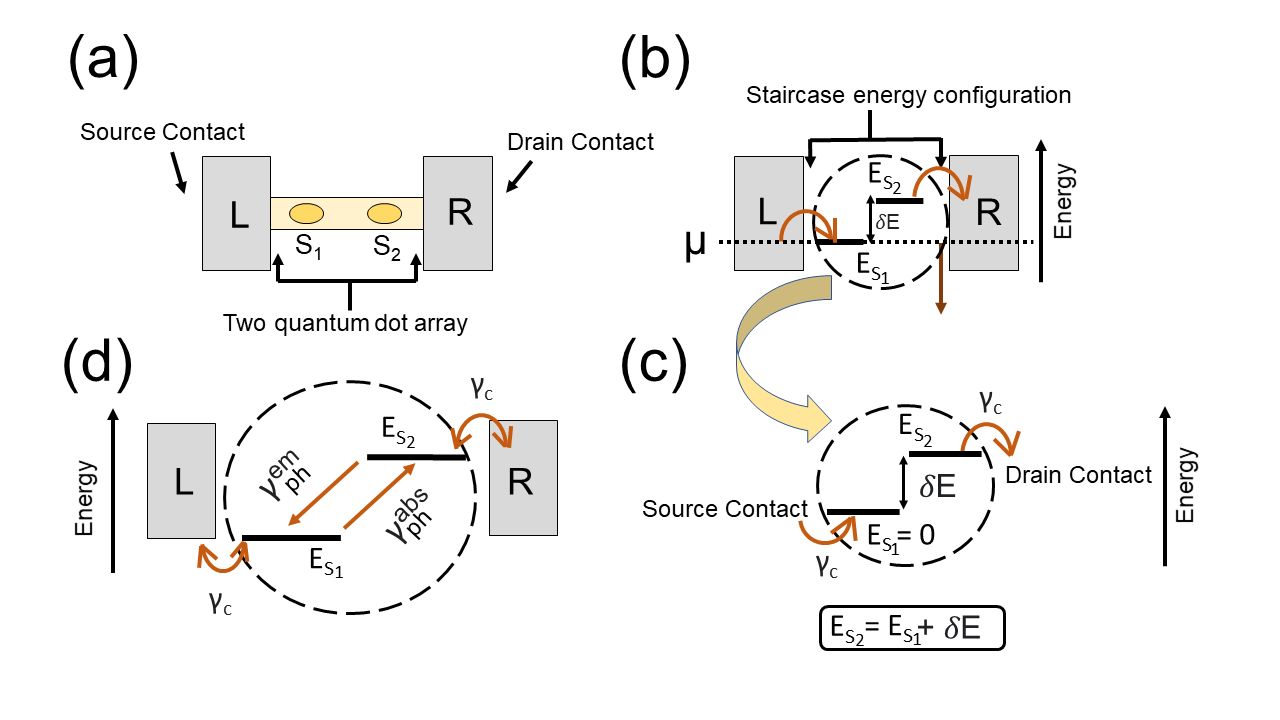}
  	\caption{Schematics of the proposed thermometer consisting of an array of two quantum dots with stair-like ground state energy configuration. (a) The proposed device is realized by embedding two quantum dots in a horizontal array, on a nanowire-like structure. (b) Illustrative diagram of the stair-like ground energy configuration of two dots. (c) A zoom into the schematics, highlighting the energy levels. The ground state energy of the first dot ($E_{S_1}$) has been assumed to be $zero$. The equilibrium Fermi level is also aligned to $E_{S_1}$. (d) An intuitive diagram representing the electron transport between the two adjacent quantum dots via energy absorption and emission. The coupling of the left (right) contacts with respective quantum dots has also been indicated.}
  	\label{fig:Fig_1}
  \end{figure*}
\begin{figure}[!htb]
	\includegraphics[width=0.5\textwidth]{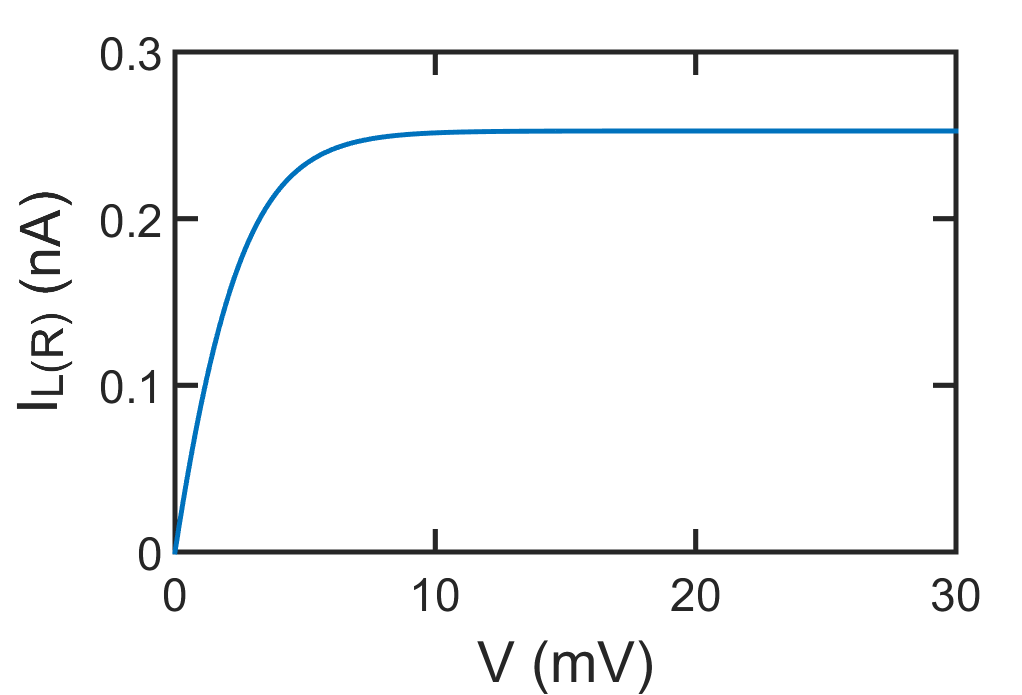}
	\caption{Variation in the reservoir-to-dot electronic current with the applied bias across the array of two quantum dots. As already mentioned to be stair-case configured, the ground states of the dots $E_{S_1}$ and $E_{S_2}$ are assumed to be separated by an energy gap of $\delta E=10k_B$ (See Fig.~\ref{fig:Fig_1}). $E_{S_1}$ is assumed to be aligned to $\mu$, and is set as reference $zero$. We choose the system temperature as $10$K.}
	\label{fig:Fig_2}
\end{figure}
\indent The set-up proposed in this paper is shown in Fig.~\ref{fig:Fig_1}, and consists of two tunnel-coupled quantum dots $S_1$ and $S_2$, which can exchange electrons with the reservoir $L$ and $R$ respectively.  The coupling between the dot $S_1 (S_2)$ and reservoir $L(R)$ is denoted by  $\gamma_c$. The dots $S_1$ and $S_2$ are tunnel-coupled to each other and share a stair-case ground state configuration with $E_{S_2}=E_{S_1}+\delta E$.  In the weak inter-dot coupling limit, a stair-case ground state  configuration with $\delta E >>\gamma_c$ suppresses electronic transport via elastic processes.   Hence, any tunnelling of electrons between the dots has to be accompanied by an absorption or emission of an energy packet $\delta E$.  A change in system temperature alters inelastic tunnelling probability between the dots, thereby modulating the current flow and exhibiting temperature sensitivity.

\section{Analysis Methodology and Transport Formulation}\label{analysis}
\indent In the setup under consideration, the rate of inelastic  absorption (emission) assisted  tunnelling  between two dots is denoted  by $\gamma^{abs(em)}_{ph}$. The  rate of elastic tunnelling ($\gamma_{el}$) is assumed negligible due to  the stair-case ground state configuration. In the typical situation of thermal equilibrium of lattice vibration, the rate of inelastic process assisted electronic tunnelling via energy absorption (emission)  is dependent on the phonon number ($N_{ph}$) associated with the energy  difference $\delta E$. Thus, the rate of inelastic tunnelling ($\gamma_{ph}^{abs(em)}$) between adjacent quantum dots can thus be written as,
 \begin{equation}
\gamma_{ph}^{abs(em)}=t_{ph}\times\left(N_{ph}+\frac{1}{2}-(+)\frac{1}{2}\right),
\end{equation}
where $t_{ph}$ is a constant proportional to the inelastic process assisted coupling between adjacent quantum dots  \cite{LijieLi, aniket_cool1}. We assume the limit of weak coupling and low-temperature to investigate the system under consideration such that  electronic transport processes via cotunneling and higher order tunneling processes, as well as via higher excited states of the quantum dots can be neglected. In the case of thermal equilibrium, the phonon number ($N_{ph}$) associated with energy $\delta E$ is given by,
	\begin{equation}
	N_{ph}=\left[exp\left(\delta E/k_BT\right)-1\right]^{-1},
	\end{equation}

    \normalsize 
To evaluate the thermometry performance, we employ the quantum master equation (QME) approach for such systems in the sequential transport regime to calculate the ground state occupancy probability   \cite{aniket_cool1}. Under thermal equilibrium of lattice vibration, the temporal dynamics  of  ground state occupancy  probability, denoted by $P_1$ and $P_2$ for the dots $S_1$ and $S_2$ respectively, can be given by the following set of equations:
\begin{align}
\frac{dP_1}{dt}=\gamma_c f_L (1-P_1) + (\gamma_{ph}^{em}+\gamma_{el}) P_2 (1-P_1)
 \nonumber \\ -(\gamma_{ph}^{abs}+\gamma_{el}) P_1 (1-P_2) - \gamma_c P_1 (1-f_L) \label{eq:probability_1}\\
\frac{dP_2}{dt}=(\gamma_{ph}^{abs}+\gamma_{el}) P_1 (1-P_2) + \gamma_c f_R (1-P_2)
 \nonumber \\ -\gamma_c P_2 (1-f_R) - (\gamma^{em}_{ph}+\gamma_{el}) P_2 (1-P_1)  \label{eq:probability_2},
\end{align} 
where $f_L$ and $f_R$ denote the occupancy probability of the reservoirs $L$ and $R$ at energy $E_{S_1}$ and $E_{S_2}$ respectively. Assuming quasi-Fermi statistics for electron occupancy at the reservoirs,
$f_{L(R)}$ can be written as
\begin{equation}
f_{L(R)}=\left(1+exp\left\{\frac{E_{S_1(S_2)}-\mu_{L(R)}}{k_BT}\right\}\right)^{-1},
\end{equation} 
where $\mu_{L(R)}$ represents the quasi-Fermi energy levels of the reservoir $L(R)$. Under the condition of quasi-equilibrium among electronic population in the reservoir $L$ and $R$, the quasi-Fermi energy in the respective reservoirs, $\mu_{L(R)}$ may be written as $\mu_{L(R)}=\mu+(-)V/2$, where $\mu$ is the equilibrium Fermi-energy through-out the entire set-up, and $V$ is the applied bias voltage with positive and negative poles connected to reservoir $R$ and $L$ respectively. Without loss of generality, the ground-state energy of the first dot $E_{S_1}$ is set as $zero$. In such a case, the quasi-Fermi functions $f_L$ and $f_R$ can be written as:
\begin{gather}
f_{L}  =\left(1+exp\left\{\frac{-\left(\mu+V/2\right)}{k_BT}\right\}\right)^{-1}  \nonumber \\
f_{R}  =\left(1+exp\left\{\frac{\delta E-\left(\mu-V/2\right)}{k_BT}\right\}\right)^{-1}
\end{gather}  
Under steady state conditions, $\frac{dP_{1(2)}}{dt}=0$. The set of differential Eqs. ~\eqref{eq:probability_1}- \eqref{eq:probability_2} are solved in steady state conditions using Newton-Raphson method to obtain the dot-occupancy probabilities. On calculation of the steady-state occupancy probabilities, the electronic current flowing out-of (into) the left (right) reservoirs can be written as:

\begin{equation}
	I_{L(R)}=\frac{2q^2}{\hbar}\left[\gamma_c \left(1-P_{1(2)}\right)f_{L(R)} - \gamma_c \left(1-P_{2(1)}\right)f_{L(R)} \right],
	\label{eq:current1}
\end{equation}
where, $I=I_L=-I_R$. In Eq.~\ref{eq:current1}, factor of 2 is introduced to account for the spin degeneracy of electrons.\\
In the aspect of nano-scale thermometry, two parameters that may be employed to gauge the performance of the proposed set-up are temperature sensitivity and efficiency (temperature sensitivity per unit power dissipation). We define the thermometer sensitivity ($\chi$) as the rate of change of electronic current between $L$ and $R$, with the system temperature $T$. \begin{equation}
\chi =  \left(\frac{dI}{dT}\right)
\end{equation}
\indent The efficiency ($\Lambda$) of the thermometer is dependent on the  sensitivity and power dissipation as:
\begin{equation}
\Lambda=\chi/P,
\end{equation}
where the power dissipation $P$  across the thermometer is defined as:
\begin{equation}
P =  V \times I,
\end{equation}
$I$ being the electronic current across the set-up and $V$ is the applied bias voltage. It should be noted that the parameter $\Lambda$ defined here is not the true efficiency related to energy conversion, as in the of heat engine. Hence, the parameter $\Lambda$ in this case is not limited to the maximum value of unity.
 
\section{Results}\label{results}
\begin{figure}[!htb]
	\includegraphics[width=0.5\textwidth]{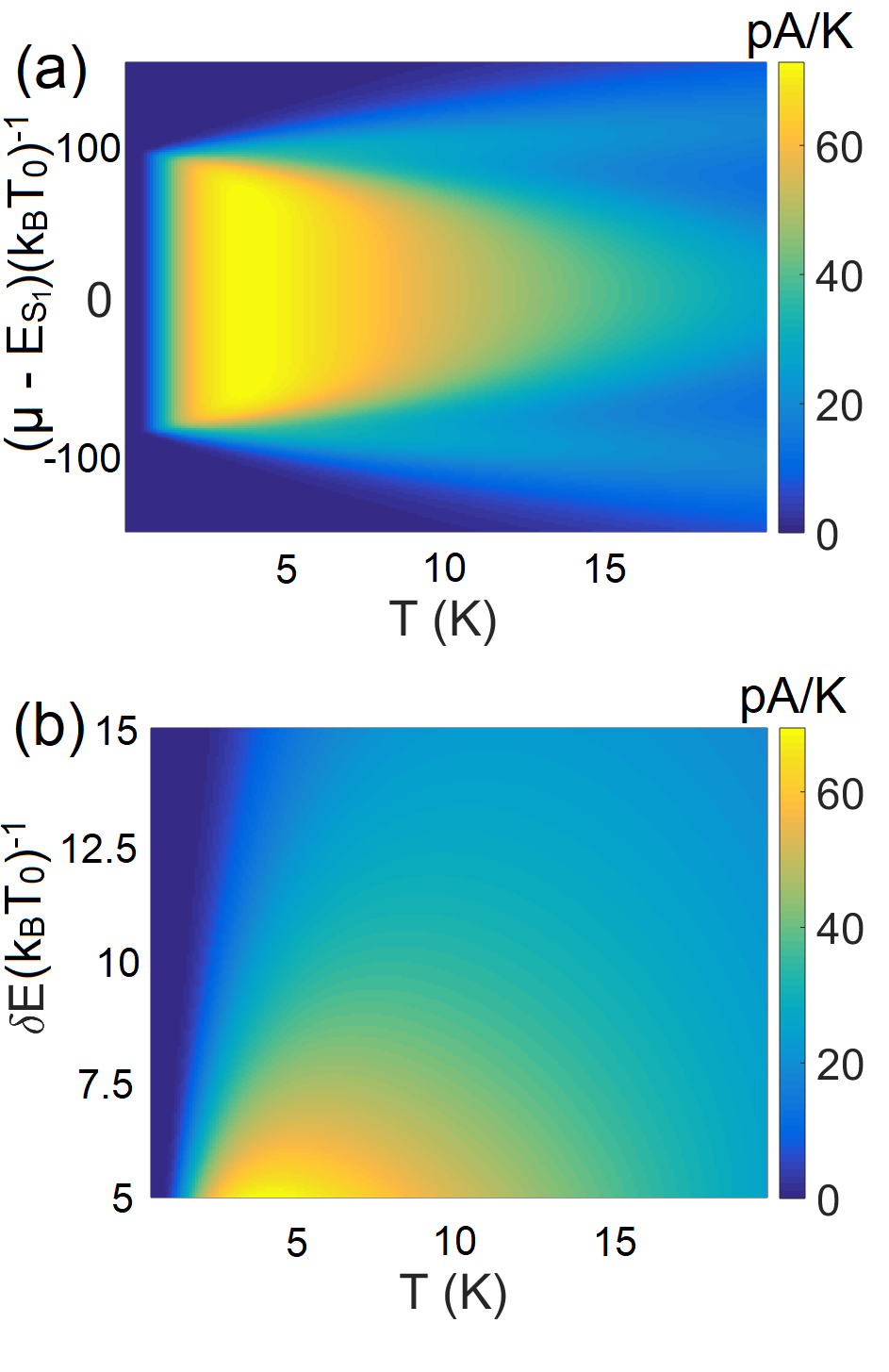}
	\caption{Colour plots demonstrating the operation regime of the proposed setup. Variation in temperature sensitivity with (a) equilibrium Fermi potential ($\mu$) and temperature $T$ (b) ground state energy difference between the two dots ($\delta E$) and temperature $T$. For both cases, the applied bias across the setup is chosen to be $V=15mV$. In the above plots, $T_0=1K$}
	\label{fig:Fig_3}
\end{figure}
\begin{figure}[!htb]
	\includegraphics[width=0.5\textwidth]{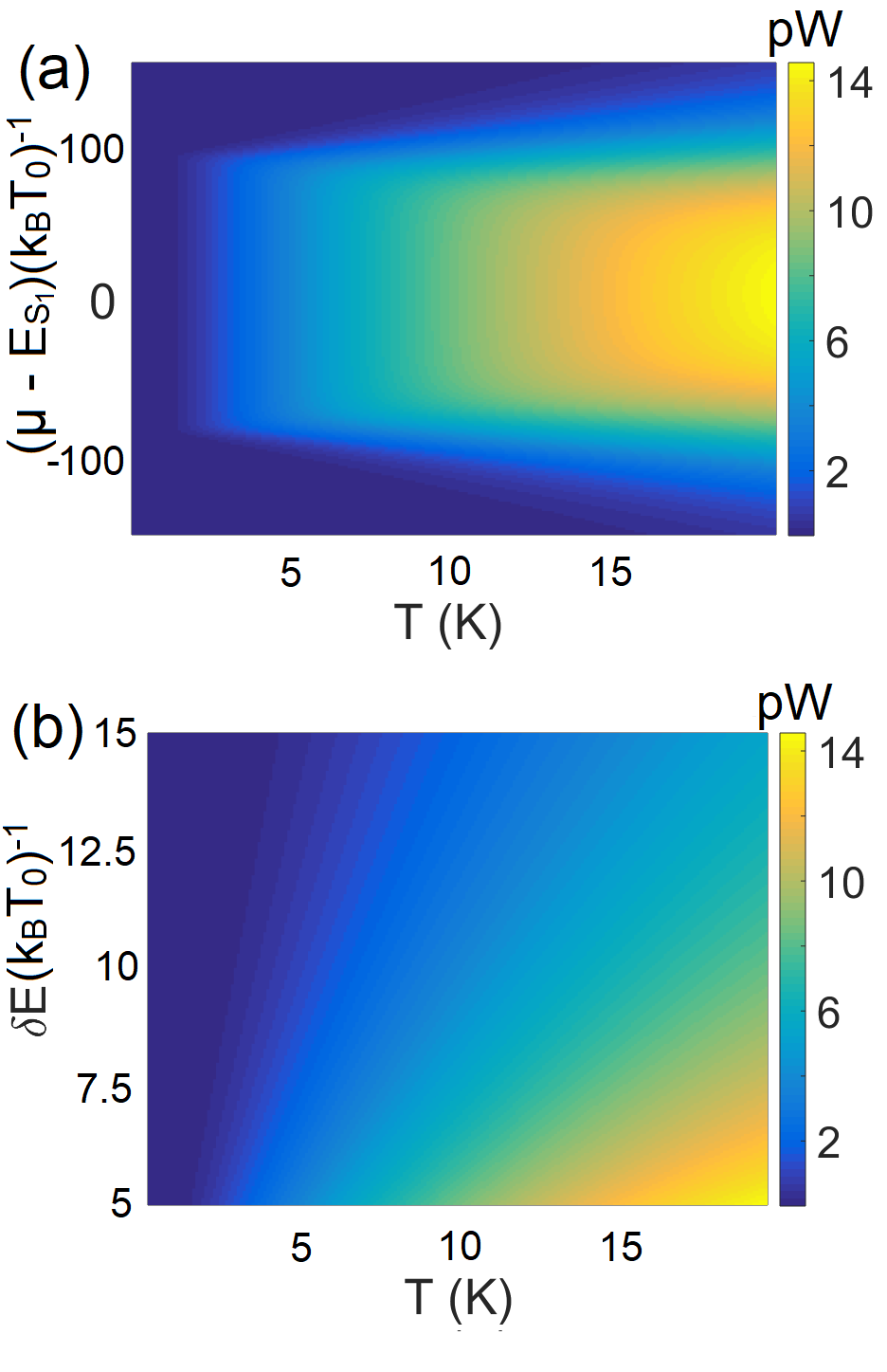}
	\caption{Colour plots demonstrating the power dissipation across the proposed setup. Variation in power dissipated with (a) equilibrium Fermi potential ($\mu$) and temperature $T$ (b) ground state energy difference between the two dots ($\delta E$) and temperature $T$. For both cases, the applied bias across the setup is chosen to be $V=15mV$. In the above plots, $T_0=1K$}
	\label{fig:Fig_4}
\end{figure}
\begin{figure}[!htb]
	\includegraphics[width=0.5\textwidth]{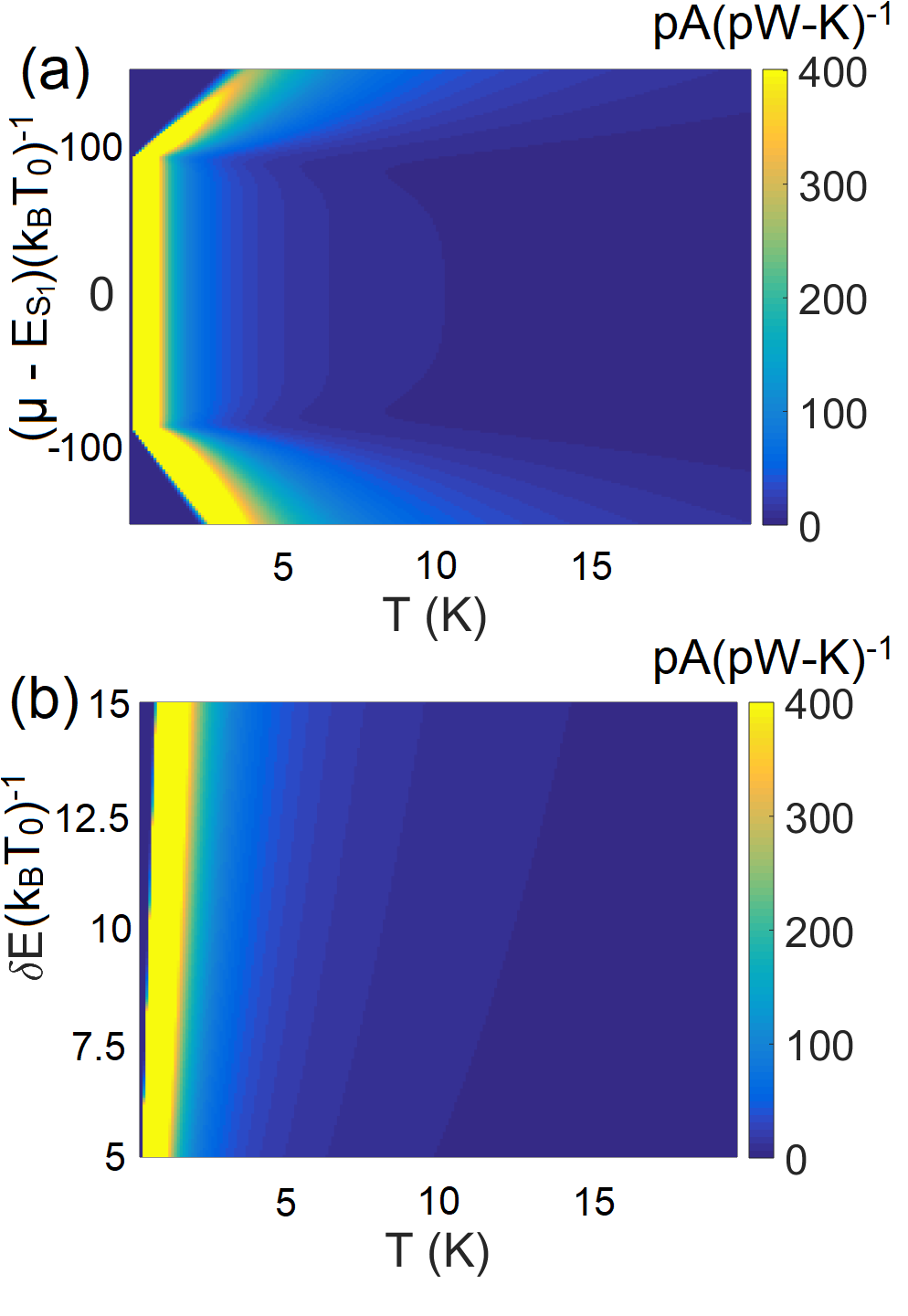}
	\caption{Colour plots demonstrating the performance of the proposed setup. Variation in  efficiency with (a) equilibrium Fermi potential ($\mu$) and temperature $T$ (b) ground state energy difference between the two dots ($\delta E$) and temperature $T$. For both cases, the applied bias across the setup is chosen to be $V=15mV$. In the above plots, $T_0=1K$}
	\label{fig:Fig_5}
\end{figure} 
\indent In this section, we  discuss the performance and the operating regime of the proposed thermometer. To investigate the proposed thermometer, we consider the limit $\delta E>>\gamma_c$, where elastic tunneling remains suppressed ($\gamma_{el}\approx 0$) due to staircase ground state configuration. Without loss of generality, we assume $\gamma_c=10\upmu$eV and $t_{ph}=1\upmu$eV. Such order of coupling parameter has been obtained from a recent experimental demonstration of non-local heat engine by Thierschmann \textit{et. al.}\cite{heatengine13}, and restrict electronic transport in the  weak coupling limit, where the rate of co-tunnelling and higher order tunnelling processes can be neglected and electronic flow can be considered to be confined in the sequential tunnelling regime. The assumption of the sequential electronic transport validates the use of quantum master equation (QME) for the analysis and performance investigation of  the proposed set-up. \\
\indent   Fig.~\ref{fig:Fig_2} depicts the variation in electronic current with the applied bias across the setup. 
To investigate the performance of the proposed set-up, we choose a bias voltage which results in the maximum saturation current. Such a choice also results in maximum possible sensitivity that can be obtained from the thermometer and  renders noise robustness  against any voltage fluctuation. For all the results that follow, the applied bias voltage is taken to be  $V=15 mV$.\\
\indent  Fig.~\ref{fig:Fig_3}(a) demonstrates the variation in sensitivity with temperature ($T$) and equilibrium Fermi potential ($\mu$), for  $\delta E= 5k_B$ ($0.43$meV). This can be explained as follows. Initially when $k_BT<<\delta E$, the phonon number $N_{ph}$ is almost zero and increases at a very slow rate resulting in very low probability of electronic tunnelling via inelastic absorption and emission. This results in low value of electronic current and hence, temperature sensitivity. With increase in temperature, the optical phonon number $N_{ph}$ gradually increases resulting in an increase in temperature sensitivity.  As $k_BT$ becomes comparable to $\delta E$, the rate of electronic tunnelling via both inelastic absorption and emission increases with $T$. The rate of increase of current depends on the temperature dynamics of $\frac{\gamma_{ph}^{abs}}{\gamma_{ph}^{em}}$. The gradual decrease, followed by saturation of the ratio $\frac{\gamma_{ph}^{abs}}{\gamma_{ph}^{em}}$ with $T$, results in the gradual decrease in sensitivity followed by saturation as $k_BT$ becomes comparable to and greater than $\delta E$.  Moreover, we note that the set-up performs optimally for a finite energy range around $\mu-E_{S_{1}}=0$. As $\mu$ deviates from $E_{s_1}$ beyond a certain limit, the temperature sensitivity gradually becomes zero due to a net decrease in electronic tunnelling between reservoirs $L$ and $R$. 
 Fig.~\ref{fig:Fig_3}(b) demonstrates the variation in sensitivity with temperature ($T$) and ground state misalignment ($\delta E$) between the adjacent  dots, assuming $\mu$ to be aligned with $E_{S_1}$. As expected, the sensitivity decreases monotonically with increase in $\delta E$. This can be explained as follows. An increase  in $\delta E$  reduces the rate of inelastic inter-dot tunnelling resulting in an overall decrease in current, and hence sensitivity. It should be noted that although the sensitivity increases with decrease in $\delta E$, a reduction in $\delta E$ beyond a certain limit might result in  the actuation of elastic inter-dot tunnelling, which results in a deterioration of sensitivity and  enhancement in dissipated power, degrading the overall performance of the set-up (see Appendix \ref{app_a}). \\
\indent Fig.~\ref{fig:Fig_4}(a) demonstrates the variation in dissipated power across the set-up with temperature $T$ and equilibrium Fermi energy $\mu$ at $\delta E= 5k_B$ ($0.43$meV).  As discussed earlier, an increase in temperature results in an enhancement in inelastic process assisted tunnelling  due to increase in phonon number ($N_{ph}$). Thus, dissipated power increases as the current increases with temperature $T$ due to an increase in electronic tunnelling rate.   Fig.~\ref{fig:Fig_4}(b) shows the variation in power dissipated across the set-up, with temperature $T$ and ground state energy difference $\delta E$ between adjacent quantum dots.  As discussed earlier,  lower values of  $\delta E$ and higher $T$ result in a higher electronic current (due to increase in phonon number $N_{ph}$) and hence, increase the dissipated power. \\
\indent Fig.~\ref{fig:Fig_5}(a) demonstrates the variation in efficiency (sensitivity per unit power dissipation) of the proposed thermometer with temperature $T$ and equilibrium Fermi energy $\mu$. It should be noted that the maximum efficiency for the proposed set-up occurs in the low temperature regime making such set-ups extremely attractive for low temperature thermometry. The system efficiency decreases with $T$ due to both decrease in sensitivity and increase in power dissipation.  Fig.~\ref{fig:Fig_5}(b) demonstrates the variation in efficiency with  temperature $T$ and ground state energy difference $\delta E$. The overall efficiency, as noted from Fig.~\ref{fig:Fig_5}(b), doesn't deviate strongly with $\delta E$.\\
  \section{Conclusion}\label{conclusion}
  To conclude, in this paper we have proposed and analysed a set-up that employs dual quantum dots with stair-case ground state configuration to achieve thermometry. The performance of the proposed set-up is actuated by an increase in inelastic absorption (emission) mediated electronic tunnelling  with temperature, resulting in a temperature sensitive conductance. It was demonstrated that the set-up offers the maximum sensitivity, as well as efficiency, in the low temperature regime, rendering it suitable for deployment in the regime of  tens of Kelvin.  In this paper, we have analyzed the set-up in the limit of weak system-to-reservoir coupling and sequential electronic transport. It would be interesting to investigate the performance as the system is gradually tuned towards the strong coupling regime. Moreover, an investigation of the effect of Coulomb blockade on the thermometry performance of the proposed set-up also constitute an interesting direction. Such explorations are left for future research. Nevertheless, the  set-up proposed in this paper may be employed to achieve high performance nano thermometers. 
  
 \indent \textbf{Acknowledgments:} Aniket Singha would like to thank financial support from Sponsored Research and Industrial Consultancy (IIT Kharagpur) via grant no. IIT/SRIC/EC/MWT/2019-20/162, Ministry of Human Resource Development (MHRD),  Government of India  via Grant No.  STARS/APR2019/PS/566/FS under STARS scheme and Science and Engineering Research Board (SERB),  Government of India  via Grant No. 
   SRG/2020/000593 under SRG scheme.\\
\appendix

	\section{Effect of elastic inter-dot tunneling on the thermometry performance} \label{app_a}
	\begin{figure*}[!htb]
	\includegraphics[width=1\textwidth]{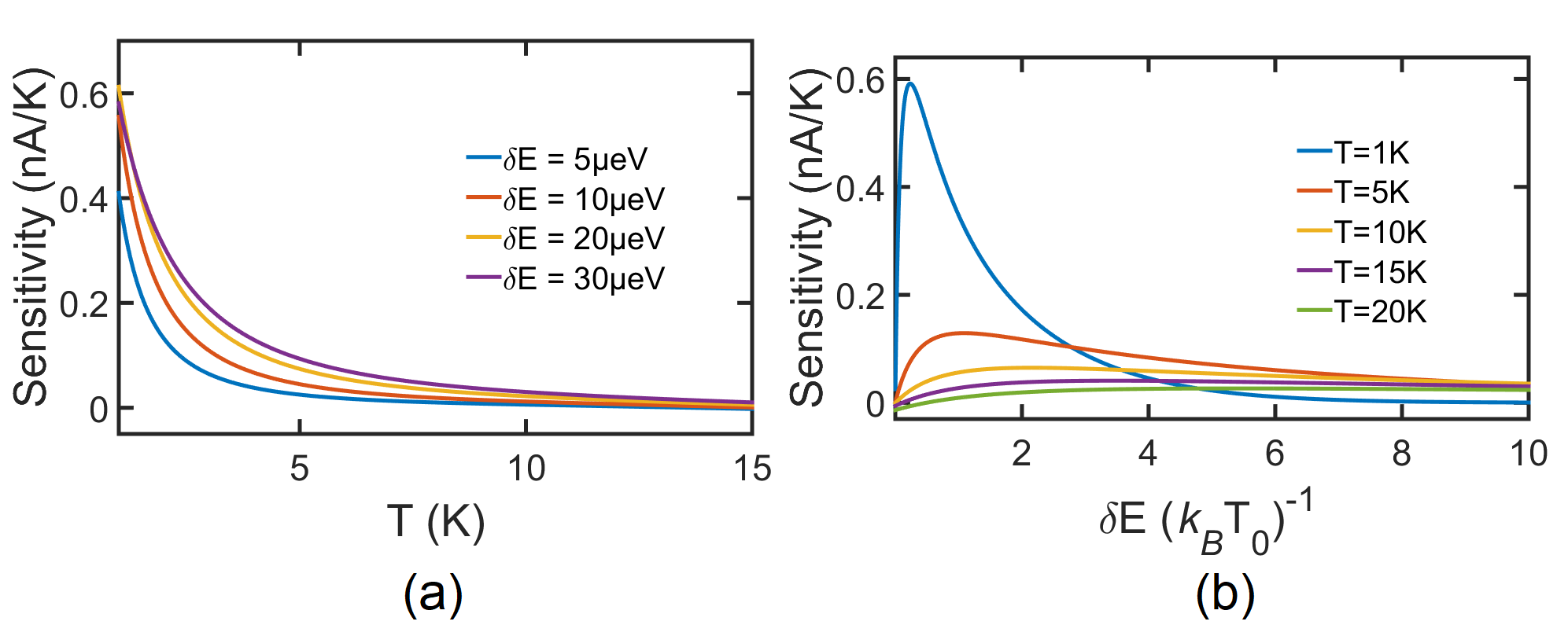}
	\caption{Variation in (a) sensitivity with temperature $T$, for various values of ground state energy difference ($\delta E$) (b) sensitivity with $\delta E$, for different temperature $T$. Here $T_0=1$K. The coupling between the left (right) contacts with respective quantum dots is assumed to be $\gamma_c=10\upmu$eV. For both cases, the voltage bias has been chosen to be $V=15$mV.}
	\label{fig:Fig_6}
\end{figure*} 
\indent Through out our discussion in the main text, we have neglected elastic tunneling processes between the two dots. Due to the stair-case ground state configuration, elastic tunnelling remains suppressed in the  limit of $\delta E>> \gamma_c$. However,  when $\delta E$ becomes comparable to or less than $\gamma_c$, due to finite lifetime-broadening of the ground states, elastic inter-dot tunneling creeps in \cite{Datta_Green,dattabook} and begins to manifest itself as a deterioration in system performance. Elastic interdot tunneling is independent of system temperature and thus the electronic current mediated via elastic processes reduces the system efficiency in addition to decreasing the temperature sensitivity (Inelastic process assisted tunelling now has to compete with elastic phenomena mediated tunneling). In the  weak interdot coupling limit, the elastic tunnelling ($\gamma_{el}$) between adjacent quantum dots can be given as \cite{singha_iopscinotes},
\begin{equation}
\gamma_{el}=t^2\times\left[\frac{2\gamma_c}{\left(\delta E\right)^2+\left(\gamma_c\right)^2}\right],
\end{equation}
where $\gamma_{el},~\gamma_{abs(em)},~\gamma_c$ have their usual meanings stated earlier in the main text and $t$ is the inter-dot hopping parameter \cite{Datta_Green,dattabook}. To investigate the impact of elastic tunneling on the system performance, we solve the set of Eqns.~\ref{eq:probability_1}-\ref{eq:probability_2}, assuming $t=1\mu eV$ and all other parameters remaining the same.\\ 
 \indent Fig.~\ref{fig:Fig_6}(a) depicts the variation in sensitivity with temperature $T$ for different values of $\delta E$ in the regime where $\delta E$ is comparable to or less than $\gamma_c$. We note that in this regime the sensitivity decreases with a decrease in $\delta E$, a trend opposite to that noted in Fig.~\ref{fig:Fig_3}  ($\delta E>>\gamma_c$). The decrease in sensitiviy with decrease in $\delta E$ occurs due to an increase the rate of elastic tunnelling ($\gamma_{el}$) which    gradually supresses inelastic process assisted tunneling  and thus reduces the system sensitivity. Fig.~\ref{fig:Fig_6}(b), illustrates the variation in sensitivity with $\delta E$, for different values of $T$. We note that the sensitivity increases as the $\delta E$ is gradually decreased till the point $\delta E$ becomes comparable to $\gamma_c$. When $\delta E$ becomes comparable to and subsequently less than $\gamma_c$, the sensitiviy decreases as elastic tunneling initiates and gradually starts dominating over inelastic process assisted tunneling.

 \bibliography{apssamp}
\end{document}